\documentstyle{mn}

\newif\ifAMStwofonts

\input epsf

\ifoldfss
  \ifCUPmtlplainloaded \else
    \NewTextAlphabet{textbfit} {cmbxti10} {}
    \NewTextAlphabet{textbfss} {cmssbx10} {}
    \NewMathAlphabet{mathbfit} {cmbxti10} {} 
    \NewMathAlphabet{mathbfss} {cmssbx10} {} 
  \fi
  \ifAMStwofonts
    \ifCUPmtlplainloaded \else
      \NewSymbolFont{upmath} {eurm10}
      \NewSymbolFont{AMSa} {msam10}
      \NewMathSymbol{\upi}     {0}{upmath}{19}
      \NewMathSymbol{\umu}     {0}{upmath}{16}
      \NewMathSymbol{\upartial}{0}{upmath}{40}
      \NewMathSymbol{\leqslant}{3}{AMSa}{36}
      \NewMathSymbol{\geqslant}{3}{AMSa}{3E}

      \let\leq=\leqslant \let\le=\leqslant
       \let\ge=\geqslant
    \fi
  \fi
\fi 

\ifnfssone
  \newmathalphabet{\mathit}
  \addtoversion{normal}{\mathit}{cmr}{m}{it}
  \addtoversion{bold}{\mathit}{cmr}{bx}{it}
  \newmathalphabet{\mathbfit} 
  \addtoversion{normal}{\mathbfit}{cmr}{bx}{it}
  \addtoversion{bold}{\mathbfit}{cmr}{bx}{it}
  \newmathalphabet{\mathbfss} 
  \addtoversion{normal}{\mathbfss}{cmss}{bx}{n}
  \addtoversion{bold}{\mathbfss}{cmss}{bx}{n}
  \ifAMStwofonts
    \ifCUPmtlplainloaded \else
      \UseAMStwoboldmath
      \usepackage{Times}
      \makeatletter
      \new@mathgroup\upmath@group
      \define@mathgroup\mv@normal\upmath@group{eur}{m}{n}
      \define@mathgroup\mv@bold\upmath@group{eur}{b}{n}
      \edef\UPM{\hexnumber\upmath@group}
      \new@mathgroup\amsa@group
      \define@mathgroup\mv@normal\amsa@group{msa}{m}{n}
      \define@mathgroup\mv@bold\amsa@group{msa}{m}{n}
      \edef\AMSa{\hexnumber\amsa@group}
      \makeatother
      \mathchardef\upi="0\UPM19
      \mathchardef\umu="0\UPM16
      \mathchardef\upartial="0\UPM40
      \mathchardef\leqslant="3\AMSa36
      \mathchardef\geqslant="3\AMSa3E

      \let\leq=\leqslant \let\le=\leqslant
       \let\ge=\geqslant
    \fi
  \fi
\fi 

\ifnfsstwo
  \DeclareMathAlphabet{\mathbfit}{OT1}{cmr}{bx}{it}
  \SetMathAlphabet\mathbfit{bold}{OT1}{cmr}{bx}{it}
  \DeclareMathAlphabet{\mathbfss}{OT1}{cmss}{bx}{n}
  \SetMathAlphabet\mathbfss{bold}{OT1}{cmss}{bx}{n}
  \ifAMStwofonts
    \ifCUPmtlplainloaded \else
      \DeclareSymbolFont{UPM}{U}{eur}{m}{n}
      \SetSymbolFont{UPM}{bold}{U}{eur}{b}{n}
      \DeclareSymbolFont{AMSa}{U}{msa}{m}{n}
      \DeclareMathSymbol{\upi}{0}{UPM}{"19}
      \DeclareMathSymbol{\umu}{0}{UPM}{"16}
      \DeclareMathSymbol{\upartial}{0}{UPM}{"40}
      \DeclareMathSymbol{\leqslant}{3}{AMSa}{"36}
      \DeclareMathSymbol{\geqslant}{3}{AMSa}{"3E}

      \let\leq=\leqslant \let\le=\leqslant
       \let\ge=\geqslant
    \fi
  \fi
\fi 

\ifCUPmtlplainloaded \else
  \ifAMStwofonts \else 
    \def\upi{\pi}
    \def\umu{\mu}
    \def\upartial{\partial}
  \fi
\fi

\title{Atlas of Tilted Accretion Discs \& Source to Negative Superhumps}
\author[M.M. Montgomery]
       {M.M. Montgomery$^{1}$\\
        $^{1}$ Department of Physics, Univeristy of Central Florida, Orlando, FL  32816, USA}
\date{Accepted .
      Received ;
      in original form 2008}

\pagerange{\pageref{firstpage}--\pageref{lastpage}}
\pubyear{2008}

\begin{document}

\maketitle

\label{firstpage}

\begin{abstract}
Using smoothed particle hydrodynamics, we numerically simulate steady state accretion discs for Cataclysmic 
Variable Dwarf Novae systems that have a secondary-to-primary mass ratio \( 0.35 \le q \le 0.55 \).  After 
these accretion discs have come to quasi-equilibrium, we rotate each disc out of the orbital plane by \( 
\delta = (1, 2, 3, 4, 5, \) or \( 20)^{o} \) to induce negative superhumps.  For accretion discs tilted 
$5^{o}$, we generate light curves and associated Fourier transforms for an atlas on negative superhumps and 
retrograde precession.  Our simulation results suggest that accretion discs need to be tilted more than 
three degrees for negative superhumps to be statistically significant.  We also show that if the disc is 
tilted enough such that the gas stream strikes a disc face, then a dense cooling ring is generated near the 
radius of impact.

In addition to the atlas, we study these artificially tilted accretion discs to find the source to negative 
superhumps.  Our results suggest that the source is additional light from innermost disc annuli, and this 
additional light waxes and wanes with the amount of gas stream overflow received as the secondary orbits.  
The nodes, where the gas stream transitions from flowing over to under the disc rim (and vice versa), precess 
in the retrograde direction.  
\end{abstract}


\section{Introduction}
Variability in non-magnetic, Cataclysmic Variable (CV), Dwarf Novae (DN) close binaries has been largely attributed to instabilities in their 
accretion discs.  Two distinct instabilities have been found to lead to two different outcomes - a thermal and viscous instability results in CV 
classical novae, recurrent novae, and DN outbursts whereas a tidal instability results in CV DN SU UMa and Nova-Like (NL) outbursts.   The former, 
normal outbursts involves cycling from low-to-high thermal and viscous states (see e.g., Meyer \& Meyer-Hofmeister 1981 and references within) and 
thus cycling from low-to-high mass transfer rates (see e.g., Lasota 2001).  The latter, larger amplitude, longer duration superoutbursts show 
characterisitc hump-shaped modulations, known as superhumps, in their light curves.  Figure 1 (Montgomery 2004) shows a CV 
family tree, and this work involves the DN and NL branches.  All classes shown in this tree are discussed in 
e.g., Smith (2006), Warner (2003), and Hellier (2001).  

\begin{figure}
\epsfxsize 3.3in
\center{\epsfbox{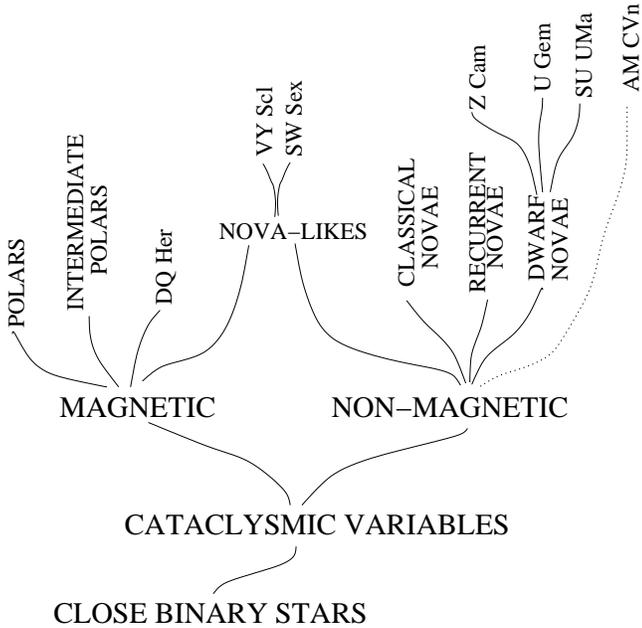}}
{\caption{Partial hierarchy of close binary systems from general (base) to more specific (top).  Mostly hydrogen systems are shown with a solid 
line and mainly helium systems are shown with a dotted line.
\label{Figure 1.}}}
\end{figure}

Observed superhump modulations in light curves are either positive if their periods are a few percent longer than the 
orbital period or negative if their periods are a few percent shorter than the orbital period.   Positive superhumps (see 
e.g., Warner 2003; Patterson 1998) are believed to result from tidal stressing of the disc by the secondary star (see 
e.g., Vogt 1982; Osaki 1985; Whitehurst 1988a,b, 1994; Hirose \& Osaki 1990; Whitehurst \& King 1991; Lubow 1991a,b; 
Murray 1996; Foulkes et al. 2004).  With the right conditions in viscosity, mass transfer rate, and secondary-to-primary 
mass ratios, the disc can outwardly expand to near the 3:1 eccentric inner Lindblad resonance (Whitehurst 1988a; Lubow 
1991a) radius where the disc becomes eccentric (e.g., Lubow 1991b; Smith et al. 2007) and is forced to change shape 
cyclically from circular to elliptical (see e.g., Simpson \& Wood 1998 and Smith et al. 2007).  The disc, however, does 
not remain stable.  Osaki (1989) attributes the larger amplitude, longer duration superoutburst to an enhanced viscous 
torque that acts on the disc once the disc becomes eccentric.  Numerical simulations show that the disc begins to slosh 
back and forth around the primary (see e.g., Lubow 1991b; Simpson \& Wood 1998), and the line of apsides of the 
oscillating disc slowly precesses in the prograde direction.  The secondary meets the line of apsides with a period that 
is slightly longer than the orbital period $P_{orb}$.  This slightly longer period is the positive superhump period 
$P_{+}$.  The positive superhump modulation in the light curve is attributed to disc dissipation from the enhanced 
spiral density waves created within the disc (Smith et al. 2007).  Both CV DN SU UMa and permanent superhump PS 
(Patterson 1999) systems show positive superhumps in their light curves.  CV DN SU UMa and PS systems have low and 
high mass transfer rates, respectively, as shown in Figure 2 (Montgomery 2004, after a figure by Osaki 1985).  The 
upper right quadrant labeled NLs refer to the magnetic NLs.  The upper left quadrant, labeled Permanent Superhumpers, 
refers to non-magnetic NLs.  Weak IPs are in between.

Unlike positive superhump theory, negative superhump theory is not as well established.  For example, a 
consensus to the source of the negative superhump has not yet been achieved.  Barrett et al. (1988) suggest 
that the source to the negative superhump is due to the varying kinectic energy of the gas stream as it 
impacts one face of a disc through a locus of points as the secondary orbits the center of mass.  Patterson et 
al. (1997) suggest that if the disc is tilted, then the gas stream can easily flow over the top and crash 
down on the inner disc and thus the source to the negative superhump is gravitational energy.  Wood, 
Montgomery, \& Simpson (2000) suggest that the source to the negative superhump is the tidal field disturbing 
the fluid flow in each of the two disc halves.  Kaygorodov et al. (2006) suggest the negative superhump may be 
due to an inner annulus bright spot that spreads due to diffusion and that is caused by matter flowing along a 
one-armed spiral density wave.  Wood \& Burke (2007) agree with Barrett et al. (1988) that the negative 
superhump modulation is from the bright spot transiting across each face of the tilted disc.  Because 
negative superhumps are found to exist alone in CV DN systems as well in systems that show both positive 
and negative superhumps and because the positive superhump modulation is suggested to come from the disc 
(Smith et al. 2007), we suggest in this work that the source to the negative superhump must also come from the 
disc and from a location that is different from that which powers the positive superhump modulation.  

\begin{figure}
\epsfxsize 3.3in
\center{\epsfbox{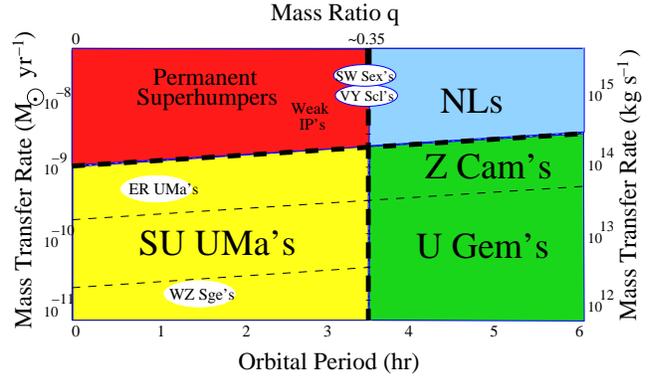}}
{\caption{Various CV systems from Figure 1 shown relative to their general orbital periods, mass ratio, and 
mass transfer rates. Dashed lines indicate approximate locations of transition zones. 
\label{Figure 2.}}}
\end{figure}

A second unknown is whether or not the accretion disc is warped or tilted.  Petterson (1977), Murray \& 
Armitage (1998), Terquem \& Papaloizou (2000), Murray et al. (2002), and Foulkes, Haswell, \& Murray (2006) 
suggest that the negative superhump modulation is related to a warped accretion disc.  Patterson et al. (1993) 
suggest that negative superhumps can be produced by a partially tilted disc.  Wood, Montgomery, \& Simpson 
(2000); Montgomery (2004); and this work show that entire disc tilts can generate negative superhumps in light 
curves.  

A third unknown is how the disc retrogradely precesses.  Both Patterson et al. (1993) and Harvey et al. (1995) 
suggest that retrograde precession in a tilted disc could be from gravitational effects similar to those on 
the Earth by the Moon and the Sun that cause the Earth to retrogradely precess (Stacey 1977).  Harvey et al. 
(1995) suggest that the secondary exerts a gravitational torque that causes the disc to slowly precess 
retrogradely but the gravitational torque will cause the tilted disc to re-align with the orbital plane.  The 
idea of a disc's retrogradely precessing line of nodes can be traced to Katz (1973), an idea that is debated in 
Kondo, Wolff, \& van Flandern (1983).  Bonnet-Bidaud, Motch, \& Mouchet (1985), continuing the debate, argue for 
a freely precessing disc and suggest this type of disc for TV Col.  This freely precessing disc model is then 
adopted by Barrett, O'Donoghue, \& Warner (1988) for TV Col, then by Patterson et al. (1993) for CVs that show 
photometric signals at the retrograde precession periods, and then by e.g., Harvey et al. (1995).  

A fourth unknown is which portion of the the disc is warped or tilted and thus which portion of the disc is in 
retrograde precession.  Either the tilted edge of a disc freely precesses due to the tidal field of the 
secondary (Katz 1973) or the secondary precesses due to the tidal field of the primary and the tilted disc 
retrogradely precesses as a consequence (Roberts 1974).  Either the inner disc is tilted, the outer disc is 
tilted, the entire disc is tilted, or the disc is warped.   Wood, Montgomery, \& Simpson (2000) and Montgomery 
(2004) show that fully tilted accretion discs precess in the retrograde direction.  

A fifth unknown is the source to disc tilt or warp.  Accretion discs have been suggested to tilt or warp via a 
portpourri of sources.  In some CVs, a disc tilt can be held constant by a gas stream fed by the 
magnetic field of the secondary (Barrett, O'Donoghue, \& Warner 1988).  A disc tilt instability can result from a 
coupling of an eccentric instability to Lindblad resonances (Lubow 1992).  A vertical resonant oscillation of the 
disc midplane can be caused by tidal interactions between a massive secondary and a coplaner primary (Lubow \& 
Pringle 1993).  A warping instability can be caused by irradiation from the primary (Pringle 1996, 1997).  A 
warping can be caused by direct tidal forces by a secondary that is orbiting on an inclined orbit.  A disc warp 
can be caused by misalignments of the spin axis of a compact and/or a magnetised primary and the disc axis.  In 
this work, we adopt no particular theory or otherwise make no statement in this work as to how the disc becomes 
misaligned with the orbital plane, saving this for future work.  We do acknowledge that Murray \& Armitage (1998) 
find that CV DN accretion discs do not seem to tilt out of the orbital plane by instabilities.  

In addition to the unknowns, observations reveal several differences between positive and negative superhump 
moduations.  Negative superhump modulations in light curves have a period that is slightly shorter than the 
orbital period, known as the negative superhump period $P_{-}$, whereas positive superhumps have a period 
slightly longer than the orbital period.  Unlike positive superhumps, negative superhumps do not seem to be 
hampered by mass ratio limits as they are found in short orbital period systems like AM CVn (see Figure 
1) and in long orbital period systems like TV Col (see NLs in Figure 2).  Negative superhumps also seem to be 
prevalent among the SW Sex and VY Scl NL systems (see Figure 2).  A third difference is the shape of the 
pulse - negative superhumps have a more equilateral triangular shape.  A fourth difference is precession - 
negative superhumps are found in systems that seem to only retrogradely precess and in systems that seem to 
simultaneously precess both progradely and retrogradely (e.g., TT Ari).  

A more complete theory on negative superhumps and retrograde precession needs to address the unknowns as 
well as the observational differences.  In addition, the theory needs to explain how one accretion disc can 
simultaneously precess both progradely and retrogradely.  Because our previous work is so successful in 
generating negative superhumps and in generating a retrogradely precessing disc, we continue to study fully 
tilted accretion discs in this work.  This work alone does not address all the unknowns and observations, but 
it does suggest a viable solution.

In this work, we provide a portion of our atlas on simulated negative superhumps and we study the source to 
the negative superhump.  For the atlas, we vary mass ratio and we maintain disc particle number, particle 
shape, viscosity, mass transfer rate, and primary mass for all simulations.  According to Paczynski (1977), a 
mass ratio \( q = M_{2}/M_{1} \leq 0.25 \) is necessary if the tidal truncation radius $R_{tides}$ is to lie 
outside the 3:1 eccentric inner Lindblad resonance radius $R_{3:1}$ for the disc to experience apsidal 
precession and for the light curve to show positive superhumps.  Patterson et al. (2005) find the 
critical ratio to be more like $ q \sim 0.3 $ based on observational studies.  Montgomery (2001), and 
references within, show that positive superhumps can be generated in systems up to a mass ratio of $q=0.33$, 
and observational results of BB Doradus (Patterson et al. 2005) tentatively agree. Because we do not wish to 
study apsidal precession and positive superhumps in this work, we limit this work to a mass ratio range \( 
0.35 \le q \le 0.55 \).  For the atlas, we tilt the disc 5$^{o}$.  In this work, we also study the source to 
the negative superhump modulation by varying the degree of disc tilt.  In \S 2 of this paper, we introduce the 
code and input parameters.  In \S 3, we suggest a source to the negative superhump modulation.  In \S 4, we 
provide a portion of our atlas on negative superhump numerical simulations, and in \S 5, we compare our 
numerical simulation results with results obtained from various observed systems.  In \S 6, we provide a 
summary and conclusions.

\section[]{The Smoothed Particle Hydrodynamics Code}
SPH is a Lagrangian method that models highly dynamical, astrophysical fluid flow as a set of interacting particles (see 
Monaghan 1992 for a review).  In SPH the local fluid properties at position $r$, a position also occupied by a 
particle, are found by sampling nearby fluid elements that are also particles and then weighting their contributions by a 
smoothing kernel.  The Monaghan \& Lattanzio (1985) interpolating smoothing kernel $W$ adopted in the Simpson (1995) code 
mimics a Gaussian but is defined to go to zero beyond a distance of 2$h$ where $h$ is the smoothing length and also the 
radius of a particle.  The code utilizes the simplest SPH form - constant and uniform $h$ and constant and uniform mass particles.  
In Montgomery (2004), to increase the number of particles in the direction normal to the disc midplane and to increase the number 
of nearest neighbors to each particle, we modify the particle size and shape from spherical to oblate spheroidal while maintaining 
the original particle volume.   Although the number of particles did increase in the $z$-direction and better packing in the disc is found, 
no appreciable improvements in results are seen.  Thus, to improve resolution, we increase particle number to 100 000 while 
maintaining spherical particles of constant radius.  To increase computational efficiency, individual particles are advanced through 
six possible timesteps, the longest of which is $\delta t_{1}=P_{orb}/200$ and the shortest of which is $\delta t_{6}=2^{-5}\delta 
t_{1}$ as described in Simpson (1995).  The choices of timesteps are determined by the particle's local environment.  In 
system units, the orbital period is normalized to $P_{orb}=2\pi$.

The code models hydrodynamics of an ideal gamma-law equation of state \( P= (\gamma-1) \rho u \) where $P$ is pressure, 
$\gamma$ is the adiabatic index, $\rho$ is density, and $u$ is specific internal energy.  The sound speed is  \( 
c_{s} = \sqrt{\gamma (\gamma -1) u}\).  The momentum and internal energy equations per unit mass are, 
respectively,

\begin{equation}
\frac{d^{2}\mathbf{r}}{dt^{2}} = -\frac{\nabla P}{\rho} + \textbf{f}_{visc} - \frac{GM_{1}}{r_{1}^{3}} \textbf{r}_{1} - \frac{GM_{2}}{r_{2}^{3}} \textbf{r}_{2},
\end{equation}

\noindent
and

\begin{equation}
\frac{du}{dt} = -\frac{P}{\rho} \nabla \cdot \textbf{v} + \epsilon_{visc}
\end{equation}

\noindent
in their most general form (Simpson 1995).  In these equations, t is time, G is the universal gravitational 
constant, ${\textbf{v}}$ is velocity, ${\textbf{f}_{visc}}$ is the viscous force, $\epsilon_{visc}$ is the energy generation due to 
viscous dissipation, and $\textbf{r}_{1} $  and $\textbf{r}_{2} $ are the displacements from  stellar masses $M_{1}$ and $M_{2}$, 
respectively.  The momentum and energy equations in SPH form for particles $i$ and $j$ are, respectively,

\begin{eqnarray}
\frac{d^{2} \mathbf{r}_{i}}{dt^{2}} & = & -\sum_{j} m_{j}\left(\frac{P_{i}}{\rho_{i}^{2}}+\frac{P_{j}}{\rho_{j}^{2}}\right)(1+\Pi_{ij})\nabla_{i}W_{ij} - \nonumber \\
                          &   & \mbox{} \frac{GM_{1}}{r_{i1}^{3}} \textbf{r}_{i1} - \frac{GM_{2}}{r_{i2}^{3}} \textbf{r}_{i2}
\end{eqnarray}

\noindent
and

\begin{equation}
\frac{du_{i}}{dt} = -\textbf{a}_{i} \cdot \textbf{v}_{i}
\end{equation}

\noindent
or if the last equation fails in preventing negative internal energies,

\begin{equation}
\frac{du_{i}}{dt} = \frac{P_{i}}{\rho_{i}^{2}} \sum_{j} m_{j} (1+\Pi_{ij})\textbf{v}_{ij} \cdot \nabla_{i} W_{ij}.
\end{equation}

\noindent
In these equations, ${\textbf{a}}$ is acceleration, $\Pi_{ij}$ is the Lattanzio et al. (1986) artificial viscosity 

\begin{equation}
\Pi_{ij}  =  \left\{ 
\begin{array}{ll}
 -\alpha \mu_{ij} + \beta \mu_{ij}^{2}	&  \qquad \mbox{$v_{ij} \cdot r_{ij}\leq 0$} \\
 0 & \qquad \mbox{otherwise}
\end{array}
\right.
\end{equation}

\noindent
where

\begin{equation}
\mu_{ij} = \frac{h v_{ij} \cdot r_{ij}}{c_{s,ij}(r_{ij}^{2} + \eta^{2})}
\end{equation}

\noindent
and $c_{s,ij} = \frac{1}{2}(c_{s,i}+c_{s,j})$ is the average sound speed, $\textbf{v}_{ij}=v_{i}-v_{j}$, $r_{ij} = 
r_{i} - r_{j}$, and \( \eta^{2} = 0.01 h^{2} \) as shown in Simpson \& Wood (1998).  

For our numerical simulations, we choose \( \alpha = \beta = 0.5 \).  Our viscosity is approximately equivalent to a 
Shakura \& Sunyaev (1973) viscosity parametrisation (\(\nu = \alpha' c_{s} H \) where $H$ is the disc scaleheight) 
$\alpha'$=0.05.  As Smak (1999) estimates $\alpha' \sim $ 0.1 - 0.2 for DN systems in high viscosity states, our 
simulations are more for quiescent systems.  As only approaching particles feel the viscous force and since neither 
radiative transfer nor magnetic fields are included in this code, all energy dissipated by the artificial viscosity is 
transferred into changing the internal energies.  As radiative cooling is not included in this code, the adiabatic index 
$\gamma$=1.01 is incorporated to prevent internal energies from becoming too large.  By summing the changes in 
the internal energies of all the particles over a specific time interval $n$, variations in the bolometric luminosity 
yield an approximate and artificially generated light curve 

\begin{equation}
L_{n} = \sum_{i} du_{i}^{n}.
\end{equation}

In all simulations, an accretion disc is generated by injecting mass from the inner Lagrange point $L_{1}$.  We inject five 
particles per major time step, 200 major time steps per orbit, or a total of 1000 particles per orbit.  This mass is 
accreted onto the primary star through an accretion disc.  The gravitational potential is treated as the sum of two point 
masses orbiting a common center of mass that is located at the origin of the coordinate system.  In system units, the 
distance between the centers of the masses is set to an orbital separation $a=1.0$.  All calculations are made in the 
inertial frame, thus eliminating Coriolis forces.  The injection velocity is based on an injection temperature $ T_{inj}$ 
= 4x10$^{4}$ K and is approximately the sound speed that has been scaled to system units.  The code adopts an approximate 
secondary mass-radius relation of Webbink (1991)  \( R_{2}=(M_{2}/M_{\odot})^{-13/15} R_{\odot} \) that applies for \( 
0.08 \le M_{2} \le 1.0 \) (Warner 2003). In this relation, $R_{2}$ and $M_{2}$ are in solar radii and solar mass, 
respectively. 

For our simulations, we assume a primary mass $M_{1}$=0.8$M_{\odot}$ and we vary the secondary mass $M_{2}$ such that 
$0.35\le q \le 0.55$.  For example, if $q=0.4$ then $M_{2}$=0.32 $M_{\odot}$ and our simulation unit length scales to \( a 
\sim 1.23 R_{\odot} \) where we have used the Webbink (1991) secondary mass-radius relationship as found in Warner (2003) 
and the Eggleton (1983) volume radius of the Roche lobe secondary

\begin{equation}
\frac{R_{2}}{a} = \frac{0.49 q^{2/3}}{0.6 q^{2/3} + ln(1 + q^{1/3})}.
\end{equation}

\noindent
The Eggleton (1983) relation is good for all mass ratios, accurate to better than 1\%.  Knowing the orbital separation, 
the orbital period could be found using Newton's version of Kepler's Third Law, but instead we adopt a Smith \& Dhillon 
(1998) secondary mass-period relation

\begin{equation}
\frac{M_{2}}{M_{\odot}} = (0.038 \pm 0.0003)P_{orb}^{1.58\pm0.09}
\end{equation}

\noindent
where $P_{orb}$ is in hours.  For $q=0.4$, the orbital period is $P_{orb} \sim$ 3.85 hr.  Taking the radius of the primary 
to be 6.9x10$^{8}$ cm, the scaled radius of the white dwarf is $R_{1}=0.0081a$.   

To obtain physical mass accretion rates, we set the particle mass at $m_{p}$=2 x $10^{14}$ kg and we assume a 
monatomic hydrogen gas, equal mass particles of equal radius $h$, and equal shape particles.  As particles are 
injected at a rate of 1000 per orbit until 100 000 particles are obtained in the disc, then the mass transfer 
rate is $ \dot{m} \sim $2 x $10^{-10} M_{\odot}$/yr for our $q=0.4$ simulation.  

A unique feature of this code is the conservation of disc particle number.  Any time a particle is accreted onto the 
secondary or primary mass or is lost from the system, a new particle is injected through $L_{1}$.  We build a disc of 100 
000 particles and maintain this number throughout the simulation as we model systems in steady state.  After building the 
disc, we artificially rotate the disc out of the orbital plane 1, 2, 3, 4, 5, or 20$^{o}$ at orbit 200 to simulate a tilted 
disc system.  We then allow the disc to evolve in the short term to a quasi-equilibrium state where particles are injected 
at the rate they are removed from the system by either being accreted onto the primary or the secondary or 
lost from the system.  The average net rate of accretion for a quasi-static disc is around 500 per orbit as 
shown in Figure 3.  As this net rate is approximately half that injected to build the disc, the steady state 
mass transfer rate reduces to $ \dot{m} \sim$ 1 x $10^{-10} M_{\odot}$/yr or a rate similar to an SU UMa in 
quiescence as shown in Figure 2.  We estimate the density of the gas near $L_{1}$ to be $\rho \sim 10^{-10}$ 
g cm$^{-3}$ using \( m_{p}=(4/3)\pi h^{3} \rho \).   As we maintain 100 000 particles in the steady state 
disc, then the total mass of the disc is  $ M_{disc} $= 100 000 x $m_{p}$ = 3.5x10$^{22}$ g or \( M_{disc} 
\sim \) 2x10$^{-11} M_{\odot}$, a negligible value compared to the mass of either star.  

\begin{figure}
\epsfxsize 2.8in
\epsfbox{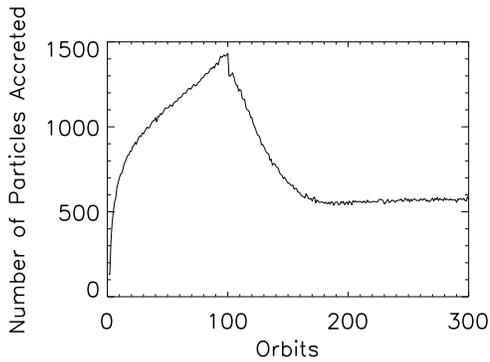}
\caption{Accretion rate as the disc is building to orbit 100 and as the disc settles into a quasi-equilibrium  
state beyond orbit 100.  The disc maintains 100 000 particles to mimic an SU UMa in quiescence and in steady 
state. }
\label{Figure 3.}
\end{figure}

If we assume the outer radius of the disc is \( R_{disc} \approx 2 r_{circ} \), or twice the circularisation radius 
(Warner 2003) where

\begin{equation}
\frac{r_{circ}}{a} = 0.0859 q^{-0.426}
\end{equation}

\noindent
and \( 0.05 \le q < 1 \), then for $q=0.4$ we find \( r_{circ} \sim 0.13 a \) and \( R_{disc} \sim 0.25 a \).   
As \( \rho << M_{disc} R_{disc}^{-3} \), self gravity is neglected in these simulations.  

To maintain a respectable number of nearest neighbors ($\ge 25$), we adjust the smoothing length so that it now depends on 
the mass ratio $q$.  Specifically, $h$ is $1/90$ of the distance between the inner Lagrange point $L_{1}$ and $M_{1}$, that 
is $h=0.011a$.  As explained in Simpson (1995), the code is three dimensional.  Each layer in the z-direction is of 
uniform thickness 2h and the entire simulation is contained within a rectangular prism that is centered at the 
origin.  The prism has sides of length 2a and depth 0.48a. Therefore, the geometrically thin disc is divided into 
approximately 22 layers.  

In Figure 4, we show for our $q=0.4$ simulation at orbit 150 the relative density as a function of both 
distance normal to the mid-plane $z$ and distance from the primary in units of orbital separation.  We also 
show the height of particles above the mid-plane and the fractional internal energy production as a function of 
distance from the primary.  In this figure, the disc has 100 000 particles and the mean number of nearest neighbors is 
64.  We note that at orbit 150, the disc has not yet been tilted.  As expected, the maximum density is along the 
midplane of the disc and peaks near the edge of the disc and the internal energy peaks in the middle of the 
disc.  The average height of the disc above the midplane as a function of distance from the primary roughly 
follows that of an isothermal axisymmetric disc to a distance $\sim$0.2a.  Beyond this distance, the 
non-axisymmetric disc and the influence of the secondary breaks the scaling.  

\begin{figure}
\epsfxsize 3.4in
\epsfbox{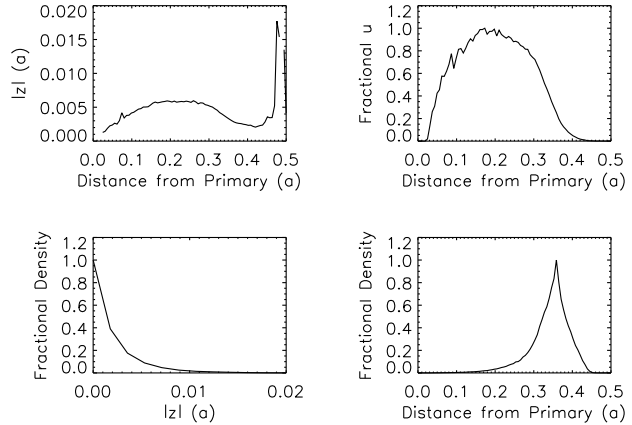}
\caption{Average height of particles above the mid-plane as a function of radius (upper left panel),  
fractional internal energy per mass as a function of radius (upper right panel), fractional density as a 
function of height z above the mid-plane (lower left panel), and fractional density as a function of 
radius (lower right panel) for our $q=0.4$ simulation at orbit 150. The disc maintains 100 000 particles. }
\label{Figure 4.}
\end{figure}

\section[]{Source to Negative Superhump Modulation}
\subsection{Catalyst to Negative Superhumps}
In Wood, Montgomery, \& Simpson (2000), we show that negative superhumps can be generated in an artificial 
light curve when the disc is tilted 5$^{o}$ out of the orbital plane.  However, we did not establish the 
minimum disc tilt that would result in statistically significant negative superhump modulations.  Once this 
minimum disc tilt is established, we can further study the role of the gas stream in the generation of 
negative superhumps.

To find the minimum disc tilt in our $q=0.4$ simulation, we stop the simulation at orbit 200, artificially rotate the disc by 
one or more degrees, and then restart the simulation.  We use a Fourier transform of the last one hundred orbits in the light 
curve as a test of whether or not negative superhumps are significantly present in the disc.  The Fourier transform of one 
hundred orbits for two, three, and four degree disc tilts is shown in Figure 5.  

\begin{figure}
\epsfxsize 3.3in
\epsfbox{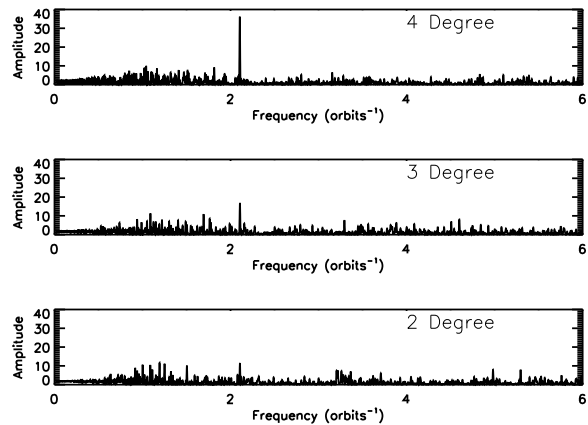}
\caption{Fourier transform of artificial light curves for $q=0.4$ simulation where the disc has been tilted two, three, or 
four degrees out of the obital plane.  Simulations are 100 orbits in duration.  Units of time are orbits, therefore 
units of frequency are orbits$^{-1}$.}
\label{Figure 5.}
\end{figure}

Although a negative superhump signal is present in all three panels, it is not significantly above the noise until the 
disc is tilted above three degrees.  In Montgomery (2004), we found for 25 000 particles simulations a similar result.  
Murray \& Armitage (1998) started with a 29 592 particle disc and did not find enough strength due to an inclination 
instability to naturally tilt the disc significantly above two degrees, even with an added one-degree tilt.  Therefore, a few 
degree tilt as a minumum seems to be in agreement for a negative superhump modulation to be stastically present.  

Note in the figure that the amplitude of the signal increases with increasing disc tilt and the growth is similar to that of a second 
order polynomial.  Wood \& Burke (2007) found that increasing the mass transfer rate also increases the amplitude and to that 
we add increasing disc tilt.  A higher amplitude signal indicates a more well-defined negative superhump modulation in the 
light curve.  For example, in Figure 6 we show orbits 220 to 230 of our $q=0.4$ simulation.  In this figure, two well-defined 
negative superhump modulations are found per orbit with maximums occurring near, e.g., orbits 220 and 230 and minimums 
occurring near, e.g., orbit 225.  The second maximum or minimum occurs approximately 1/2 orbit earlier or later in each 
example.  

\begin{figure}
\epsfxsize 3.5in
\center{\epsfbox{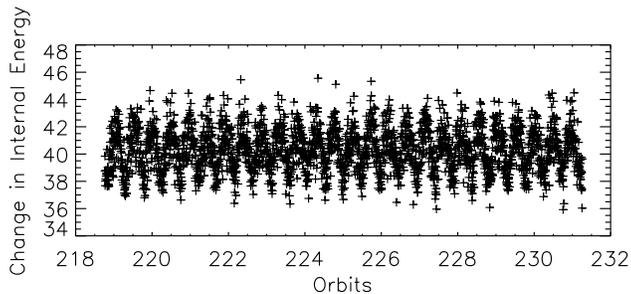}}
\caption{Light curve from an accretion disc that has been tilted 20$^{o}$ out of the orbital plane at orbit 200.  In this simulation, 
$q=0.4$.  Two negative superhump modulations are shown per orbit. }
\label{Figure 6.}
\end{figure}

To generate this figure, we tilted the disc 20$^{o}$ at orbit 200 such that the line of nodes is perpendicular to 
the line connecting the primary and secondary center of masses, herein referred to as the line of centers.  Therefore the line 
of centers is parallel to the disc's line of antinodes. As the secondary orbits the center of mass in the counter clockwise 
direction, the secondary travels roughly one quarter of an orbit when it encounters the disc's line of nodes.  Traveling nearly 
another quarter orbit, the line of centers is once again perpendicular to the line of nodes and parallel to the disc's line of 
antinodes and the secondary encounters a disc face but this time the secondary is facing the other side of the disc.  After 
traveling nearly three quarters of an orbit, the secondary once again meets the disc's line of nodes, and after traveling 
nearly one full orbit, nearly to orbit 201, the secondary once again encounters the disc's line of antinodes.  At this time 
the secondary is facing the same side of the disc that the secondary encountered at orbit 200. During this one 
orbit, the gas stream interacted with the disc edge twice at the bright spot (i.e., at the disc's line of 
nodes) and twice with the disc face (i.e., at the disc's line of antinodes).  The cycle continues as the 
secondary orbits with the gas stream flowing over the disc for nearly one-half orbit and then transition to 
flowing under the disc for nearly another one-half orbit.

Because the negative superhump modulation is well defined in this high disc-tilt simulation, the varying kinetic energy of 
the gas stream as it impacts the face of the disc through a locus of points as the secondary orbits (Barrett et al. 
1988) may be the source to the negative superhump.  To test this idea, we find where the gas stream strikes the disc when 
the disc is tilted a minimum disc tilt that leads to statistically significant negative superhump modulations (i.e., 4$^{o}$).  
In Figure 7, we show disc side-views for our $q=0.4$ simulation.  We compare untilted and tilted discs.  In both panels, 
the secondary is not shown but we show the gas stream leaving the inner Lagrange point and striking the disc.  In 
the top panel, we show orbit 200 (O200) of an untilted disc and thus the accretion stream is striking the disc 
at the bright spot as expected.  Some of the gas particles flow both above and below the disc after striking 
the disc rim.  In the bottom panel, we show orbit 220 (O220) when the disc is emitting its maximum negative 
superhump modulated light (refer to Figure 6).  As shown, the gas stream is striking the lower half of the disc 
edge yet still at the bright spot, and more of the gas particles flow under the disc than over the disc.  As 
negative superhumps are statistically signficant from a disc tilted 4$^{o}$, we conclude that the gas stream 
does not have to impact the face of the disc in a locus of points to generate negative superhumps.  We do 
acknowledge that negative superhumps cannot be present without the gas stream. In Figure 7, the aspect ratio 
is exaggerated to emphasize the role of the gas stream.

\begin{figure}
\epsfxsize 3.in
\center{\epsfbox{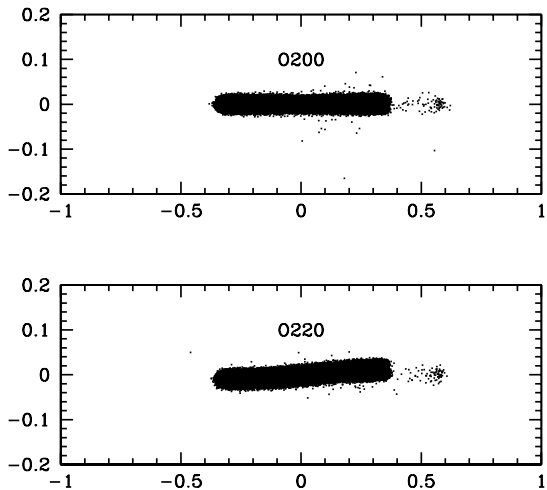}}
\caption{Edge-on views of the gas stream striking the bright spot of a non-tilted disc (upper panel) and a disc that has 
been tilted 4$^{o}$ at orbit 200 (lower panel).  The upper panel is the $q=0.4$ simulation at orbit 200 before the disc is 
artificially tilted and the lower panel is the simulation at a later time, orbit 220.  In both panels, the aspect ratio is 
exaggerated to emphasize the role of the gas stream in the generation of negative superhumps from a minimally tilted disc.}
\label{Figure 7.}
\end{figure}

As a disc tilt of 4$^{o}$ can induce a negative superhump signal well above the noise yet the flared shape of the disc causes 
the gas stream to mostly strike at the rim of the disc as opposed to cleanly flowing over the disc edge, our results suggest 
that the bright spot does not necessarily have to transit the face of disc to induce negative superhumps as suggested by 
others.  However, the gas stream that strikes the disc rim and flows mostly over one face of the disc per half orbit does 
appear to be involved.  If the disc is not tilted a few degrees, then the gas stream cannot mostly overflow one face of the 
disc and negative superhumps are not present.  Therefore, as expected, disc tilt of a few degrees is a major catalyst to 
induce negative superhumps, and without the gas stream, negative superhumps would not be present.  

\subsection{Location of Negative Superhump Light Source} 
To identify when and where the negative superhump occurs within the disc, we need to find the orbital parameters 
and the negative superhump phase.  From the Fourier transform of orbits 200-300 in our $q=0.4$ numerical 
simulation that has been tilted 5$^{o}$ about the primary at orbit 200, we identify the negative superhump 
period to be \( P_{-}=0.95 \) orbits.  To convert this period into real units, multiply one orbit by the orbital 
period \( P_{orb} \sim 3.85 \) hours.  Knowing these two periods, we can determine the nodal precessional period 
by \( P_{Pn}^{-1} = P_{-}^{-1} - P_{orb}^{-1} \) to be \( P_{Pn} = 72.2 \) hours or 19 orbits.  Another relation 
that can be found is the nodal superhump period excess \( \epsilon_{n} = 1 - \frac{P_{-}}{P_{orb}} = 0.049 \).  
Having identified the negative superhump period to be \( P_{-} = 3.66 \) hours, we can fold the last 100 orbits 
on this period to find the negative superhump phase.

In Figure 8, we show folded superhump lightcurves from orbits 200-300 for our $q=0.4$ numerical simulation.  
Having found two negative superhump modulations per orbit from Figure 6, we expect to find two negative 
superhump pulse shapes per orbit in Figure 8 as our light curves are essentially bolometric, independent of 
viewing angle, or similar to having integrated over $4\pi$ sr.  Observers typically see only one face of a 
tilted disc and hence only one superhump modulation per orbit.  As shown in this figure, the maximum superhump 
light occurs approximately before the half-orbit phase ( \( \phi \sim 0.4 \) ) and full-orbit phase ( \( \phi 
\sim 0.9 \) ) and the minimum occurs just before the quarter-orbit and three quarter-orbit phases.  Notice one 
pulse amplitude is higher than its half-orbit counterpart.  

\begin{figure}
\epsfxsize 3.in
\center{\epsfbox{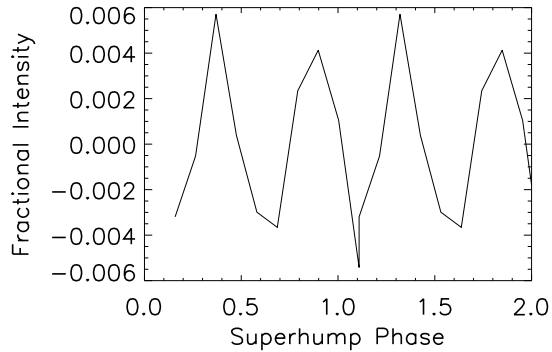}}
\caption{Folded bolometric light curve of orbits 200-300 for our q=0.4 simulation where the disc has been 
tilted 5$^{o}$ out of the orbital plane.}
\label{Figure 8.}
\end{figure}

Knowing the negative superhump maximum phases, we now seek to identify when the disc is emitting this extra 
light.  From Figure 6, we identified orbits 220 and 230 as examples of approximate maximums and orbit 225 as 
an example of an approximate minimum for a disc that has been tilted 20$^{o}$.  As the $q=0.4$ tilted disc 
precesses every 19 orbits, other near maximums should occur near orbits 201 and 210 and other near minimums should 
occur near orbits 206 and 215.  With 200 frames per orbit and maximums and minimums occurring nearly every 
half orbit as shown in Figure 6, maximums should occur near frames 0, 100, and 200 of orbits 210, 220, 
and 230.  Likewise, minimums should occur near frames 0, 100, and 200 of orbits 215 and 225.  Minimums should 
also occur near frames 50 and 150 of orbits 210, 220, and 230.  Likewise, maximums should also occur near 
frames 50 and 150 of orbits 215 and 225.  As shown in Figure 7, a maximum occurs when the secondary is nearly 
in line with the disc's line of antinodes and a minimum occurs when the secondary is nearly in line 
with the disc's line of nodes, in agreement with theory.  

Knowing the approximate orbit and frame when the additional light is being emitted, we now search to identify 
that portion of the disc that is contributing mostly to the negative superhump modulation. As the disc 
contributes to most of the light from CV systems, the source to the negative superhump modulation should be 
from the disc itself, or a portion thereof.  As the negative superhump modulation has a higher amplitude and 
becomes more coherent with increasing disc tilt, we plot in Figure 9 the sum of each particle's change in 
internal energy for frame 0 orbit 215 (faint line, lower panel), frame 0 orbit 220 (bold line, lower panel), 
frame 150 of orbit 220 (faint line, upper panel), and frame 200 of orbit 220 (bold line, upper panel) from a 
disc that has been tilted 20$^{o}$.  The dashed line in both panels is from an untilted disc at frame 0 of 
orbit 150 like that shown in the upper right panel of Figure 4.  

As shown in Figure 9, the internal energy distribution from a untilted disc is vastly different than that from a tilted 
disc.  In an untilted disc, the number of particles around 0.1a and 0.35a is $\sim$600 per bin with each bin of 0.0125a 
equal width.  The distribution of the total energy produced resembles a bell curve that is centered near 0.2a 
with $\sim$825 particles contributing to the maximum light.  By orbit 215 frame 0, after the disc has been 
tilted 20$^{o}$ at orbit 200, the number of particles per bin around 0.1a and 0.35a have increased to 
$\sim$1200 and $\sim$900, respectively.  The distribution of energy production has changed from that of a bell 
curve to a more linear, negatively sloped curve with more particles emitting more light near 0.1a.  Thus the 
inner disc now contributes to most of the light emitted from tilted discs, as expected.  A second difference between the 
untilted and tilted disc fractional light emitted is the notch in the tilted disc curves near 0.27a.  The notch suggests 
that a ring has formed near 0.27a that emits less light than from adjacent annuli, and the source of the cooling ring 
appears to be from the gas stream striking the disc face.  

\begin{figure}
\epsfxsize 3.3in
\epsfbox{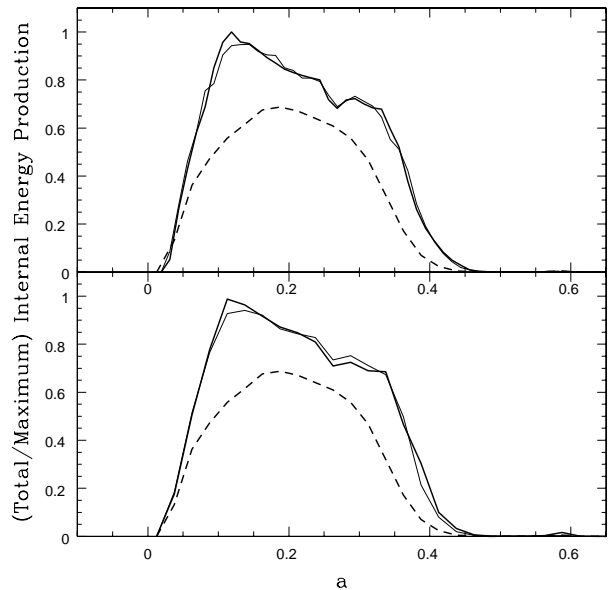}
\caption{The total change in internal energy is shown as a function of distance from the primary that is located at $a=0$ 
for frame 150 of orbit 220 (faint line, upper panel), frame 200 of orbit 220 (bold line, upper panel), frame 0 of orbit 215 
(faint line, lower panel), frame 0 of orbit 220 (bold line, lower panel) of a disc that has been tilted 20$^{o}$.  Frame 0 
of orbit 150 for an untilted disc is shown in both panels as dashed lines.  The radial distance of each particle from the 
primary is determined from the Pythagorean Theorem.  
}
\label{Figure 9.}
\end{figure}

\begin{figure}
\epsfxsize=2.3in
\centerline{\epsfbox{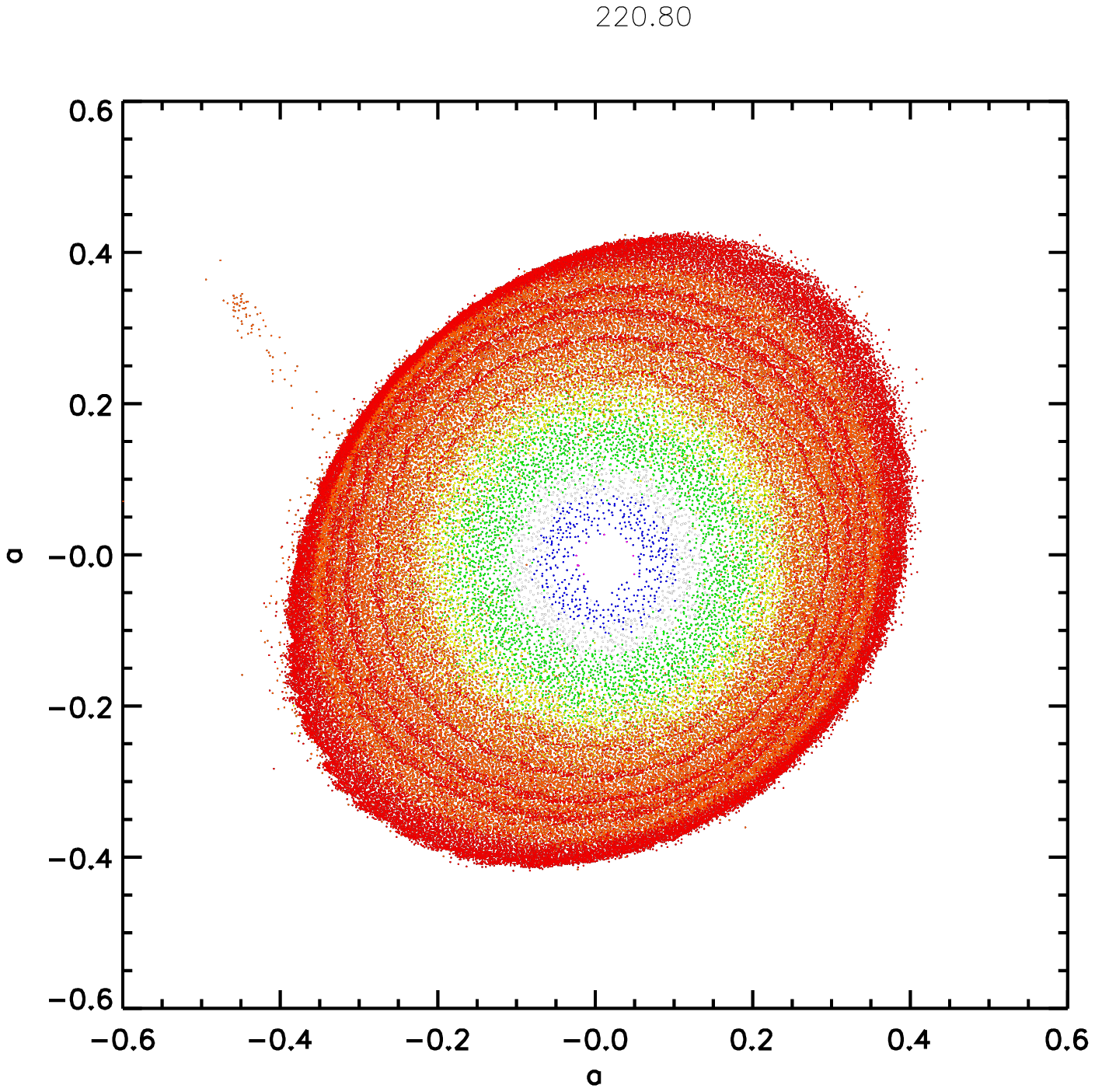}}
\bigskip
\epsfxsize=2.3in
\centerline{\epsfbox{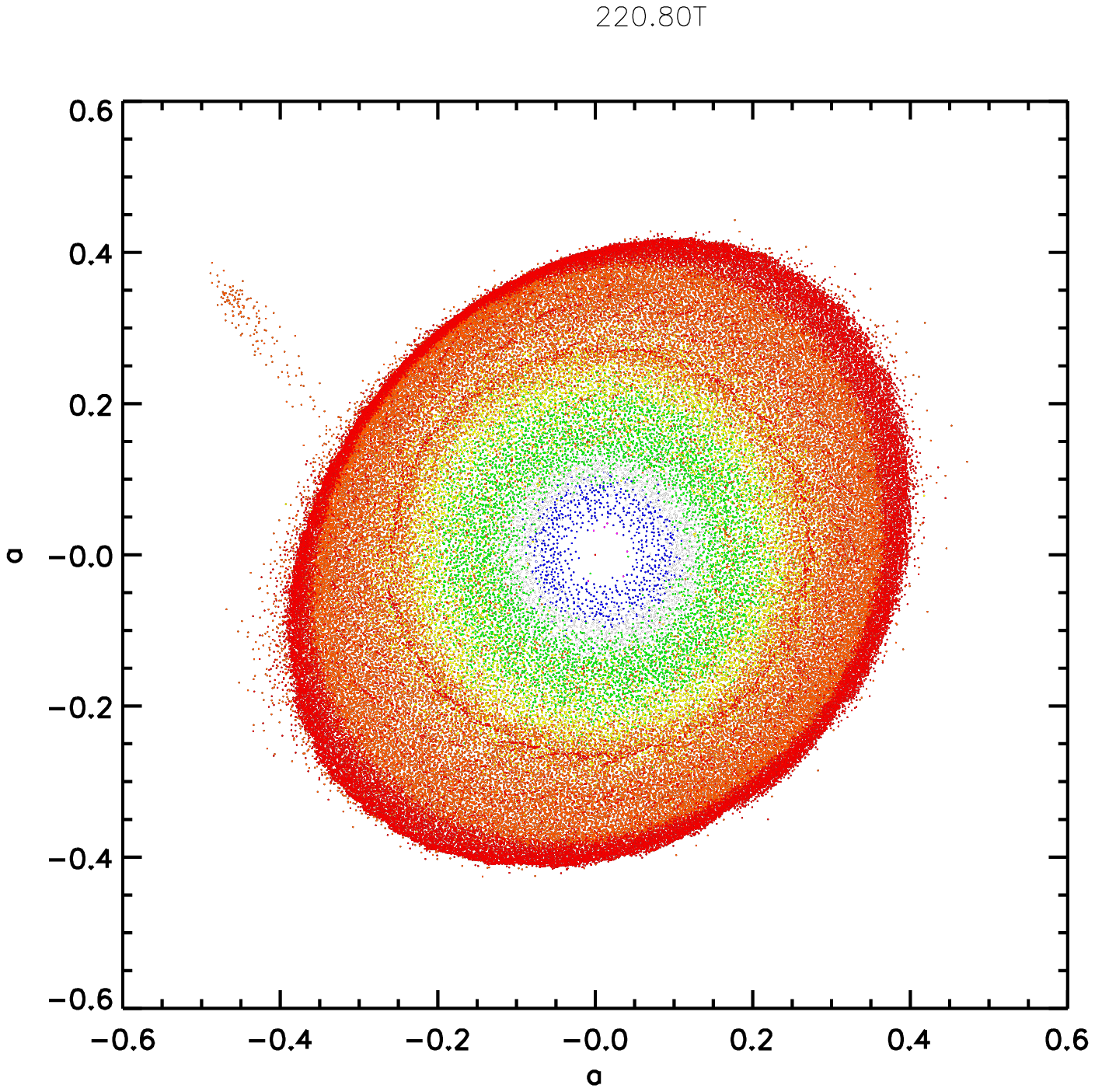}}
\caption{Face-on disc snapshots for q=0.4, orbit 220, frame 80 of a 20$^{o}$ tilted (bottom frame) and untilted 
(top frame) disc.  The gas stream is the colour orange.  The colour red represents less energetic particles whereas blue 
represents more energetic.  Relative internal energies are orange/red $\sim$ 0.6\%, yellow/red $\sim$ 4\%, green/red $\sim$ 
8\%, white/red $\sim$ 18\%, blue/red $\sim$ 29\%, violet/red $\sim$ 83\%.   
}
\label{Figure 10.}
\end{figure}

The largest change between faint and bold lines in the figure occurs near 0.1a, and thus this location appears 
to be the source to the negative superhump modulation.  Comparing the curves in Figure 9, the disc is alternating between 
nearly minimum states (faint lines - frame 0 of orbit 215 and frame 150 of orbit 220) and nearly maximum states (bold 
lines - frames 0 and 200 of orbit 220) nearly twice per orbit and this cycle continues through to orbit 300.  When the 
disc is in a maximum state, more light is being emitted near 0.1a than when the disc is in a minimum state.  With 
approximately 150 particles per bin and three to four bins, $\sim$500 particles near $\sim$0.1a are contributing to the 
additional light when the disc is near a maximum state (bold lines) than when the disc is near a minimum state (faint 
lines).  This number of particles is approximately half the injection rate per orbit for this numerical simulation:  
Approximately 850 particles are lost to the primary and $\sim$100 to the system each orbit, and thus $\sim$950 particles 
are injected per orbit to maintain 100 000 particles in the disc.  Both the number of particles contributing to 
the negative superhump modulation and the number of particles injected per orbit are few compared to the number 
of particles within the disc.  However inner annuli are more energetic than outer annuli, and thus more light 
is emitted per particle from inner annuli.  We see from a comparison of frames 0 of orbits 215 and 220 and from 
frame 150 and frame 200 of orbit 220 that the light emitted from the disc has increased $\sim$5\% between 
0.1a-0.15a.  If we assume a 5\% net increase in disc light, then from \( m_{1} - m_{2} \) = 2.5 log \( 
\frac{L_{2}}{L_{1}} \) we find a 0.053 decrease in apparent magnitude which is within the observed range of 
0.03-0.6 mag.  

Although we know that the additional negative superhump light occurs near a radius $r=0.1$a, we have not yet 
identified whether the emission is from a partial or an entire ring.  Therefore, we show in Figure 10 face-on discs 
for our q=0.4 simulation.  Both panels show orbit 220, frame 80 (additional colour plots in relative density and 
in relative energy emitted can be found at www.exoplanets.physics.ucf.edu/$\sim$montgomery).  The panels show 
energy emitted per particle for a disc tilted 20$^{o}$ (bottom panel, labeled 220.80T) and for an untilted disc 
(top panel, labeled 220.80).  The colour red represents least energetic whereas the colour blue represents more 
energetic.  The gas stream is the colour orange and, therefore, the disc rim is emitting less energy than the gas 
stream.  Notice a red cooling ring appears near radius r$\sim$0.27a, the approximate location where the gas 
stream strikes the face of the tilted disc.  The additional particles emitting more energy from outer annuli, the 
cooling ring, and the additional particles emitting more energy from inner annuli of the tilted disc are in 
agreement with line curves shown in Figure 9.  

In the untilted disc (top panel, Figure 10), annuli with radii r $>$ 0.2a alternate colours red and orange whereas annuli 
with radii r $<$ 0.2a gradually increase in energy released with decreasing radius.  In comparison, the tilted disc 
appears more blended in colour.  Annuli with radii r $>$ 0.2a are nearly all at the same energy as that emitted 
by the gas stream.  That is, the alternating red and orange ring pattern found in the outer annuli of the 
untilted disc is nearly lost in the tilted disc due to the removal of the gas stream striking the disc rim at the 
bright spot.  On the other hand, annuli with radii r $<$ 0.2a in the tilted disc appear more blended because of 
the increased gas particle migration from outer annuli (e.g., the yellow ring blending with the green ring).  In 
addition, when the disc is tilted 20$^{o}$, the gas stream strikes the disc face and some of the gas particles skip off 
the face of the disc towards inner annuli as shown by the additional orange coloured particles in the inner annuli of the 
tilted disc.  Both the additional particle migration and the additional ring of particles that skip off the disc face to 
inner, more energetic annuli appear to contribute to negative superhump maximum.  When the disc is at a negative superhump 
minimum, the gas stream is striking the bright spot at the disc rim like that shown in the top panel of Figure 10. Hence 
the additional orange coloured particles that skip off the disc face and the additional particle migration to inner annuli 
that occur when the disc is at a negative superhump maximum are not present.

Our results suggest that as gas particles strike inner, less dense, more energetic annuli, the number of particles 
increase in that annuli and the disc responds to the increase in number by producing more light.  The combination of 
additional particles in inner annuli and the additional light that each particle emits via disc tilt is suggested by our 
work to be the source of the negative superhump modulation.  Wood \& Burke (2007) show this inner annuli brightening over 
one half orbit although they attribute the additional light to the transit of the bright spot across the face of the disc.  
Kunze et al. (2001) find inner annuli brightening from their studies of various rates of mass transfer overflowing disc 
rims, also supporting these results.

\subsection{Disc Configuration for Negative Superhumps}
Having shown that the disc has to be tilted for the gas stream to overflow a disc face and having identified that the gas 
particles need to reach the inner, more energetic annuli for the negative superhump modulation to occur, we seek to 
establish a viable disc configuration that can answer the previously identified unknowns and observations.  Others have 
suggested that the outer disc is tilted, the inner disc is tilted, or the disc is warped.  We suggest a fully tilted disc.  

If the disc is fully tilted about the line of nodes, then the gas stream can overflow one disc face and reach inner, more 
energetic annuli.  As the secondary orbits, each face of the disc can release this additional light nearly once per orbit.  
Observers would see only one negative superhump modulation if the optically thick disc is more face-on to the observer.  A 
full disc tilt also agrees with the findings that the amplitude of the negative superhump increses with increasing disc 
tilt or with increasing mass transfer rate.  If the disc is tilted higher, then the more gas particles can reach inner 
annuli.  If instead the disc tilt is low and the mass tranfer rate is high, then more particles can plow through to inner 
annuli.  

A full disc tilt configuration can explain the observed negative superhump periods and nodal percessional periods.  
It can explain the disc's retrograde precession of the line of nodes, that is the location within the disc where the gas 
stream transitions from flowing over to flowing under the disc and vice versa.  This configuration explains the location 
within the disc that powers the negative superhump modulation. As this location is different than that which powers the 
positive superhump modulation, this configuration can explain how both modulations can occur simultaneously within the 
same disc.  Although this work does not address all of the unknowns and observations discussed in the Introduction, it does 
provide a viable disc configuration that addresses many in the list.  A full disc tilt configuration looks promising.

\section[]{NUMERICAL SIMULATION RESULTS}
We present in this work a portion of our atlas that focuses on negative superhumps only and thus we present 
simulations for mass ratios 0.35$\le$ q $\le$0.55.  Figure 11 shows simulated light curves for these various 
negative superhump numerical simulations.  In all panels of this figure, the disc is created up through orbit 
100.  After the disc has settled in the short term, the disc is artificially tilted 5$^{o}$ at orbit 200.  In all 
simulations, we see a small rise in internal energy production at orbit 200.  This rise is due to the disc tilt.  
The most significant result is the lack of features from orbits 200 to 300 in all simulations.  With a mass transfer 
rate like that of an SU UMa or a U Gem (see Figure 2) and with a 5$^{o}$ tilt, the negative superhump 
modulation is hard to see.

\begin{figure}
\epsfxsize 3.3in
\epsfbox{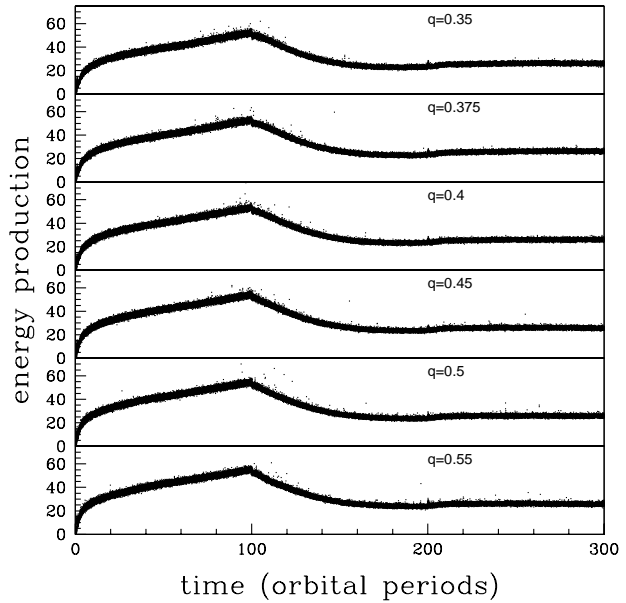}
\caption{Light curves from accretion discs that have been tilted 5$^{o}$ out of the orbital plane at orbit 200.  Mass 
ratios are shown.  Units of orbital periods are orbits. }
\label{Figure 11.}
\end{figure}

Figure 12 shows the Fourier transforms to the light curves in Figure 11.  In this figure, the negative 
superhump signal is found at a frequency greater than two because two superhump modulations are created in our 
simulations, one per half orbit, because the secondary encounters the first line of nodes at a negative 
superhump period that is less than one half the orbital period and because we effectively integrate over 4$\pi$ 
sr to determine the light emitted over time.  No harmonics of statistical significance are generated at the 
mass transfer rate used in these simulations.  In simulations with 25 000 particles (Montgomery 2004) injected 
at a rate of 2000 particles per orbit, twice that of these simulations, harmonics are generated.  Therefore, 
mass transfer rate is identified to be a source to changing mode frequencies.  Also shown in this figure is an 
increasing amplitude with mass ratio, the strongest of which is from a disc with the smallest radius.  Therefore 
we can identify mass ratio as a source to increasing the amplitude of the negative superhump modulation in 
addition to disc tilt and mass transfer rate. 

\begin{figure}
\epsfxsize 3.3in
\epsfbox{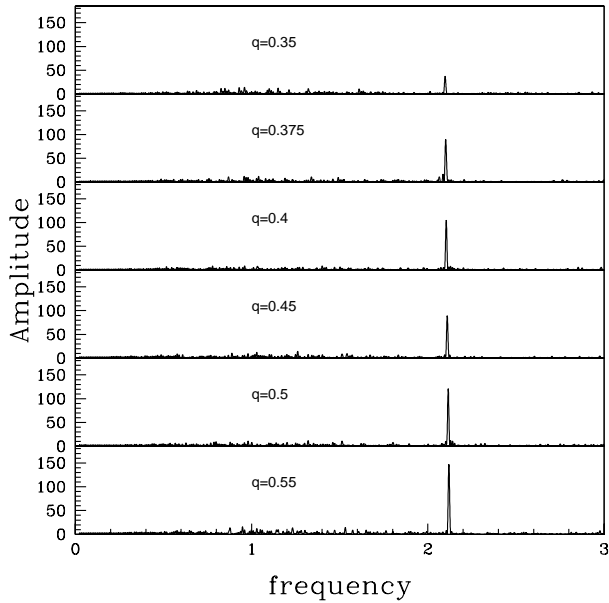}
\caption{Fourier transforms of light curves in Figure 11.  Mass ratios are shown.  Units are orbit$^{-1}$.}
\label{Figure 12.}
\end{figure}

A non-linear least squares fit to each Fourier transform is computed and the standard deviation to twice the 
negative superhump frequency $2\nu_{-}$ is found.  This data is listed in Table 1 along with the calculated 
negative superhump period, its propagated error, and the nodal superhump period excess.  The units for period 
are hours whereas the units for frequency are orbits$^{-1}$ to be consistent with the Fourier transform 
figure.  To convert to real units, multiply one orbit by the orbital period.  

Figure 13 shows the negative superhump phases for two orbits of our $q=0.55$ simulation.  We show the $q=0.55$ 
pulse shape as a representative sample because of the strong signal in the Fourier transform and because of 
the uniform pulse shapes.  To create the pulse shape, the $q=0.55$ light curve in Figure 11 is folded on the 
corresponding negative superhump period given in Table 1 for orbits 200-300.  Comparing Figures 8 and 13, we find that 
negative superhump pulse shapes look similar regardless of mass ratio.  We also find that a higher mass ratio results in 
more similar amplitudes of the $\phi$ = 0.4 and $\phi$=0.9 phases.  These results suggest that the light emitted from each 
face of the disc twice per orbit becomes more equal with increasing mass ratio and thus smaller disc size.  

Also of interest is the shape of the pulses as they resemble equilateral triangles like those found from 
observations.  The shape suggests that the modulated light uniformly increases to a maximum and then uniformly 
decreases to a minimum.  As inner annuli supply the extra light, this result suggests that particles within 
the inner annuli wax and wane their internal energies uniformly as well.  As the gas stream that overflows the 
disc feeds the inner annuli and because we keep the gas stream mass transfer rate fairly constant in our 
simulations, our results suggest that the equilateral shaped negative superhumps obtained from observations 
may involve a time-independent gas stream mass transfer rate.

\section[]{COMPARISONS WITH OBSERVATIONAL DATA AND DISCUSSION}
Table 2 lists observational data including orbital periods, negative superhump periods, and nodal period excesses for several 
cataclysmic variable (CV) systems.  If errors are known, they are listed in parentheses.  We select long orbital period 
systems so that we may compare their observational results with our higher mass ratio numerical simulation results.
In the table, NL is novalike, PS is permanent superhumper, and IP is intermediate polar.  

\begin{table*}
 \centering
 \begin{minipage}{200mm}
  \caption{Negative Superhump Simulation Data}
  \begin{tabular}{@{}lllllll@{}}
  		&\multicolumn{2}{c}{Frequency ($P_{orb}^{-1}$)}& \multicolumn{3}{c}{Period (hr)}&\\
   		\cline{2-3} 
		\cline{4-6}
q & $2\nu_{-}$ & $\sigma_{2\nu_{-}}$ & $P_{orb}$ & $P_{-}$ & $\sigma_{P_{-}}$  & $\epsilon_{n}$ \\
 \hline
0.35   & 2.098 & 0.004 & 3.540 & 3.374 & 0.006 & 0.0467 \\
0.375  & 2.105 & 0.004 & 3.698 & 3.514 & 0.006 & 0.0498 \\
0.4    & 2.104 & 0.004 & 3.852 & 3.662 & 0.006 & 0.0494 \\
0.45   & 2.110 & 0.004 & 4.150 & 3.934 & 0.007 & 0.0521 \\
0.5    & 2.115 & 0.003 & 4.436 & 4.195 & 0.007 & 0.0544 \\
0.55   & 2.119 & 0.003 & 4.712 & 4.447 & 0.007 & 0.0562 \\
\hline
\end{tabular}
\end{minipage}
\end{table*}

\begin{table*}
 \centering
 \begin{minipage}{180mm}
  \caption{Negative Superhump Observational Data and Calculated Processional Periods and Nodal Superhump Period Excesses}
  \begin{tabular}{@{}lllllll@{}}
   System & Type &q  & $P_{orb}(d)$& $P_{p_{n}}(d)$  &  $P_{n}(d)$  & $\epsilon_{n}$ \\
 \hline
 TV Col$^{a,b}$     & IP          &$0.4\pm0.05^{c}$                     & 0.2292    & 3.973 & 0.2167    & 0.0545 \\
 V751 Cyg$^{d}$     & NL, VY Scul &                                     & 0.1445(2) & 3.806 & 0.1394(1) & 0.0353(2)\\
 PX And$^{e,f}$     & NL, SW Sex  & $0.329(11)^{f}$                     & 0.14635(1)& 4.228 & 0.1415    & 0.0331 \\
 BH Lyn$^{f,g}$     & NL, SW Sex  &$0.45+0.15-0.10^{h}$                 & 0.15575(1)& 3.413 & 0.1490    & 0.0433 \\
                    &             &0.301(15)$^{f}$                      &           &       &           &       \\
                    &             &0.41$\pm$0.26$^{g}$                  &           &       &           &       \\
 KR Aur$^{i}$       & NL, VY Scul &$0.60^{j}$                           & 0.1628    & 4.489 & 0.1571(2) & 0.0350(2) \\
 AT Cnc$^{k}$       & Z Cam, PS   &$0.32-1.04$                          & 0.2011(6) & 5.051 -&0.1934(8) -&0.0179(10) -\\ 
                    &             &                                     &           & 11.03 & 0.1975(8) & 0.0383(10) \\
 TX Col$^{l}$       & IP          &                                     & 0.2375    & 1.669 & 0.2083    & 0.1229 \\
 SDSS J040714.78-064425.1$^{m}$&SU UMa?&                                & 0.17017(3)& 6.727 & 0.166(1)  & 0.024(1) \\
 HS 1813+6122$^{n}$ & NL, SW Sex  &                                     & 0.1479    & 3.166 & 0.1413    & 0.0446 \\
 V2574 Oph$^{o}$    & Nova        &                                     & 0.14773   & 3.429 & 0.14164   & 0.0412 \\ 
 RX 1643+34$^{p}$   & NL, SW Sex  &                                     & 0.120560(14)& 3.917&0.11696(8)& 0.03000(8) \\
\hline
\end{tabular}
\\
  $^{a}$Hellier \& Buckley (1993), $^{b}$Retter et al. (2003), $^{c}$Cropper et al. (1998), $^{d}$Patterson et al. (2001), \\
  $^{e}$Patterson (1998), $^{f}$Patterson et al. (2005), $^{g}$Dhillon et al. (1992), $^{h}$Hoard \& Szkody (1997), \\ 
  $^{i}$Kozhevnikov (2007), $^{j}$Kato et al. (2002), $^{k}$Kozhevnikov (2004), $^{l}$Retter et al. (2005), $^{m}$Ak et al. (2005), \\ 
  $^{n}$Rodriguez-Gil P. et al. (2007), $^{o}$Kang et al. (2006), $^{p}$ Patterson et al. (2002)\\
\end{minipage}
\end{table*}

\begin{figure}
\epsfxsize 3.in
\center{\epsfbox{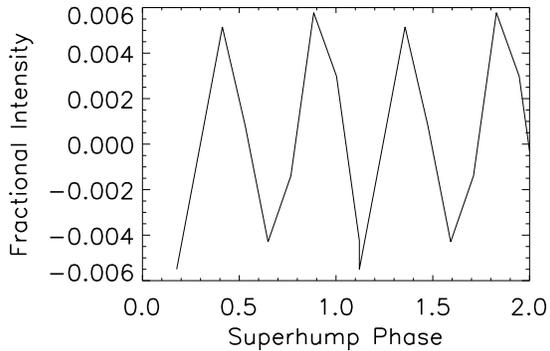}}
\caption{Folded bolometric light curve of orbits 200-300 for $q=0.55$ simulation where the disc has been 
tilted 5$^{o}$ out of the orbital plane at orbit 200.  The unit of the superhump phase is orbits.}
\label{Figure 13.}
\end{figure}

From a quick scan of Table 2, the nodal period excess for TX Col does not follow the trend set by the other 
observational data.  TX Col's orbital period is the longest of the set, longer than TV Col's, and its negative 
superhump period is not as long.  Thus, further observations of this object is suggested and we do not 
consider this object any more in this study.  The tentatively assigned candidacy type of SU UMa for the 
Sloan object does not agree with its observational period (see Figure 2) and an observational revisit to this 
object may also be needed but we do consider this object in our study.

Of the mass ratios that are known, all are higher and nearly all have high mass transfer rates as implied by their 
subclasses (see Figure 2).  Excluding the Sloan object, the observational data suggests that negative superhumps are 
more common in high mass transfer systems than in low mass transfer systems.  Also, negative superhumps do not seem to 
be a feature of the U Gem class indicating that U Gem's may be immune to disc tilt.  

Figure 14 shows a comparison of the observational nodal period excess (*) to the numerical nodal period excess 
($\triangle$) values  listed in Tables 1 and 2, respectively.  The values are compared relative to their 
orbital periods and errors for orbital periods can be taken as the maximum width of each symbol.  Of the 
observations, HS 1813+6122 agrees reasonably well with our numerical simulation.  Since we simulated a red 
dwarf secondary using the Smith \& Dhillion (1998) secondary mass-period relation, this system may have a red 
dwarf secondary as it is within the period gap.  As the observational data do not suggest a trend, 
a secondary mass-period relation other than the Smith \& Dhillon (1998) utilized in this work may need to be 
studied for these long orbital period systems.  

\begin{figure}
\epsfysize 2.5in
\center{\epsfbox{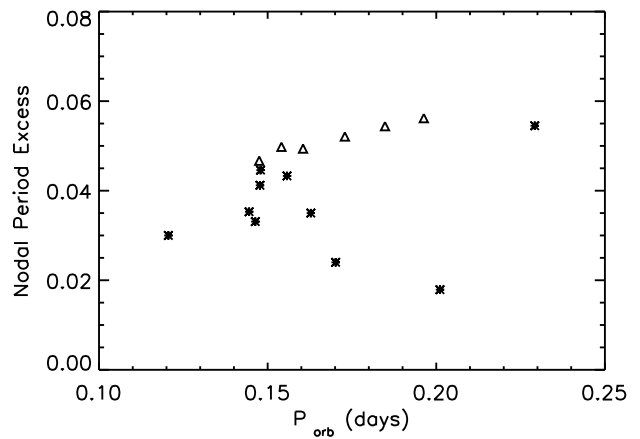}}
\caption{Nodal superhump period excess as a function of orbital period.  Observational data are shown as stars and numerical data are 
shown as triangles. Data are from Tables 1 and 2.}
\label{Figure 14.}
\end{figure}

\section{SUMMARY, CONCLUSIONS, AND FUTURE WORK}
We artificially tilt SPH numerical simulated accretion discs out of the orbital plane by various degrees and for 
various mass ratios, and we establish their negative superhump periods, orbital periods, nodal precessional periods, 
and nodal superhump period excesses.  We confirm our earlier work that tilted discs do wobble and lead to a 
retrograde precession of the line of nodes.  In addition to increased mass transfer rates, we find that 
increasing disc tilt and higher mass ratios result in higher amplitude and more coherent negative superhump 
modulations.  Simulations of higher mass ratios also show more similar amplitude pulses.

We find that discs need to be tilted a few degrees for the negative superhump to be statistically well above 
the noise.  Discs that are tilted at least a few degrees allow the gas stream to overflow the disc rim and 
reach inner, less dense, more energetic annuli.  These inner annuli wax and wane in emission as the secondary 
orbits.  Thus, the innermost annuli are identified to be the location that powers the negative superhump modulation, 
a location that is different from that which powers the positive superhump.  We find that the minimum disc 
tilt that results in negative superhump modulations of statistical importance is 4$^{o}$.  However, because the 
gas stream still strikes the disc rim at this low disc tilt, our results suggest that the bright spot does not 
necessarily have to transit the face of the disc for negative superhump modulations to be generated.  If the 
disc tilt is high, however, then the gas stream strikes the disc face and causes a dense, cooling ring to be 
generated within the disc.  The total light emitted as a function of radius has a notch near the location of 
the gas stream disc interface.  

We find that an accretion disc that is entirely tilted around the line of nodes can be a viable disc 
configuration for systems that show only negative superhump modulations or that show both negative and positive 
superhump modulations.  This disc configuration can explain the retrograde precession of the disc's line of 
nodes, that is the location of where the gas stream transitions from mostly flowing over the disc to mostly 
flowing under the disc and vice versa.  It can also explain how a gas stream can reach inner annuli that powers 
the negative superhump.  It can also explain how two modulations can co-exist in the same disc as the modulations occur 
in different locations within the disc.  The apsidal superhump light comes from spiral density waves that extend radially 
through the disc and we suggest that the negative superhump light comes from inner disc annuli.  Future work involves 
numerical simulations showing how both modulations can occur simultaneously in the same disc.  This work also 
does not explain how accretion discs become tilted, and we save this for future work.  

The approximately equilateral triangular shape of the superhump pulse suggests that this additional brightening waxes and 
wanes uniformly with the constant increase or decrease of gas particles that reach inner annuli of the disc.  That is, the 
uniform shape of the pulse suggests that the mass transfer rate is time-independent. Our results also show that lower mass 
transfer rates do not generate harmonics.  

Negative superhumps are found in observational systems with high mass ratios, high orbital periods, and high mass 
transfer rates.  The U Gem class of CVs seems to be immune to tilted discs.  This work does not study SU UMas, and 
thus we cannot say much further on this class in this study.  As no trend is found in negative superhump period 
excess of observational systems that have long orbital periods, additional secondary mass-period relations other than 
those studied in this work are suggested.  

\section*{Acknowledgments}
The author would like to acknowledge Alon Retter for his useful comments from many years ago as well as Matt 
Wood.  We would like to thank the UCF/UF Space Research Initiative and the AAS Small Research Grant that enabled us 
to generate this work.  Many thanks to our undergraduate research student Mark Guasch whose work on colouring the changes 
in internal energy and density in face-on and edge-on disc plots for this study, several of which can be found at our 
website exoplanets.physics.ucf.edu/$\sim$montgomery.  Lastly, we would like to thank the anonymous referee.

\label{lastpage}

\end{document}